\documentstyle[aps,epsf,prc]{revtex}
\begin{document}
\title{Determination of the $g_{\rho\sigma\gamma}$ coupling constant through the process $\gamma+N\to N+V$ with circularly polarized photons}
\author{Michail P. Rekalo \footnote{ Permanent address:
\it National Science Center KFTI, 310108 Kharkov, Ukraine}
}
\address{Middle East Technical University,
Physics Department, Ankara 06531, Turkey}
\author{Egle Tomasi-Gustafsson}
\address{\it DAPNIA/SPhN, CEA/Saclay, 91191 Gif-sur-Yvette Cedex,
France}
\date{\today}

\maketitle
\begin{abstract}
We suggest a method to determine the product of the coupling constants $g_{\rho\sigma\gamma}g_
{\sigma NN}$, in absolute value and relative sign, through the $t$-dependence of the two possible T-even asymmetries for the photoproduction of vector mesons, $\gamma+p\to p+V^0$, $V^0=\rho$ or $\omega$, in the collision of circularly polarized photons with a polarized proton target.  In the framework of a $\pi+\sigma$-model, these asymmetries are very sensitive to the relative role of $\pi$- and $\sigma$-exchanges, in the near threshold region. 
\end{abstract}

\section{Introduction}
The $\rho$- and $\omega$-meson photoproduction, $\gamma+N\to \rho(\omega)+N$, in the near threshold region, $E_{\gamma}\le$ 2.5 GeV, can be described through quite simple mechanisms: the $\pi$-exchange for $\omega$ photoproduction
and the scalar $\sigma$-exchange for $\rho$ photoproduction. The two coupling constants, $g_{\pi NN}$ and $g_{\omega\pi\gamma}$, which enter in the calculation of the absolute values of the differential cross section for $\gamma+p\to p+\rho^0$, are well known. The existing experimental data on the $t-$dependence of the differential cross sections in the near threshold region can be reproduced 
with the help of few parameters such as the cut-off parameters of phenomenological form factors \cite{Fr96,Zh98,Ti99,La02,Ba02}.
 
In the case of $\rho^0$-photoproduction, $\gamma+p\to p+\rho^0$, the contribution of $\pi$-exchange is too small, and the suggested mechanism of $\sigma$-exchange depends on the product of two unknown coupling constants
$g_{\rho\sigma\gamma}g_{\sigma NN}$, which can be fitted in order to obtain an agreement of such model with the experimental cross section $d\sigma(\gamma p\to \rho^0p)/dt$. As a result, using the constant $g_{\sigma NN}$, from $NN$-potential \cite{Ma89}, it was possible to find the numerical value for the unknown coupling constant $g_{\rho\sigma\gamma}\simeq 3$ (for the standard choice of the corresponding Lagrangian). The determination of this coupling constant in the framework of QCD sum rules, gives also relatively large values: $g_{\rho\sigma\gamma}=3.2\pm 0.2$ \cite{Go01} and $g_{\rho\sigma\gamma}=2.2\pm 0.4$ \cite{Al02}. The analysis of the $ \rho^0\to \pi^+\pi^-\gamma$ decay in the framework of two possible mechanisms the $\sigma$-exchange, $\rho^0\to \sigma^0+\gamma \to \gamma+\pi^+ +\pi^-$ and brehmstrahlung radiation by charged pions produced led to even larger values of 
$g_{\rho\sigma\gamma}$ \cite{Go00}. In the last case, it was not possible to find an unique value for this constant, because the final result depends on the mass and the width of the $\sigma$-meson, which value is given in a broad range,  0.4 GeV$\le m _{\sigma}\le $1.4 GeV, 0.6 GeV$\le \Gamma _{\sigma}\le $1 GeV \cite{PdG}. 

Note, in this connection, that a large value of the $g_{\rho\sigma\gamma}$ coupling constant has been considered also for the explanation \cite{Mi00} of the HERMES
effect \cite{Ac99}, concerning the anomalous nuclear dependence of the cross section for the inclusive electroproduction on nuclei at relatively small momentum transfer squared, $Q^2$: the increasing absorption of virtual photons with longitudinal polarization and the decreasing cross section with transversally polarized photons.

For completeness we should note that the $g_{\omega\sigma\gamma}$ coupling constant (with the corresponding electromagnetic form factor for the space-like momentum transfer) is important for the discussion of very actual problems, concerning the meson exchange currents in the physics of the electromagnetic deuteron structure \cite{Ch74,Hu90,Vo95}. 

Finally let us mention that the $V\sigma\gamma$-coupling constant can be estimated on the basis of the existing experimental data on the radiative decays of $\rho^0$- and $\omega$-mesons:
\begin{equation}
\omega(\rho^0)\to \pi^0\pi^0\gamma.
\label{eq:eq1}
\end{equation}
The very small branching ratios of these decays ($Br\simeq 10^{-5}$, \cite{Ac00}) are in contradiction with large values of the $g_{V\sigma\gamma}$-coupling constant, which have been derived before, using theoretical considerations (QCD-sum rules) \cite{Go01,Al02}, and experimental data on $\gamma+p\to p+\rho^0$ and the HERMES effect \cite{Mi00}. In the last two  cases, the $g_{V\sigma\gamma}$-coupling constants have been determined by fitting the data, in the framework of a model. In the case of the process $\gamma+p\to p+\rho^0$,  only the $\sigma$-contribution has been taken into account \cite{Fr96}. But in the near threshold region, other mechanisms contribute: the nucleon exchanges in the s- and t-channel, the excitation of various nucleonic resonances and even extrapolation of Pomeron exchange to the threshold region. The same is correct for the model of the HERMES effect. Only the decays (\ref{eq:eq1}) can be considered as direct methods to determine the $g_{V\sigma\gamma}$-coupling constants. 

In this paper we plan to discuss another possible method to determine the $g_{V\sigma\gamma}$-coupling constant -through the measurement of T-even asymmetries in the collision of circularly polarized photons with a polarized proton target. The circular polarization of the photon results from the brehmstrahlung of longitudinally polarized electrons. Such beams are available at MAMI (Mainz) \cite{Au97} and at the Jefferson Laboratory, where it has recently been used in the study of the deuteron photodisintegration $\vec\gamma+d \to n+\vec p$-with the measurement of the polarization of the outgoing proton \cite{Wi01}. In principle such experiments, as $\vec \gamma+\vec p\to p+V^0$ or $\vec \gamma+p\to \vec p+V^0$, $V^0=\rho^0$ or $\omega$, are actually feasible.
\section{Asymmetries of $\vec \gamma+\vec p\to p+V^0$-processes}
The dependence of the differential cross section of the process  $\vec \gamma+\vec p\to p+V^0$ on the polarizations of the colliding particles can be written in the following form \cite{Re02}:

\begin{equation}
\displaystyle\frac{d\sigma}{d\Omega}(\vec\gamma\vec p)=\left ( \displaystyle\frac{d\sigma}{d\Omega}\right)_0
\left (1+\lambda t_x{\cal A}_x+\lambda t_z{\cal A}_z\right ),
\label{eq:sig}
\end{equation}
where $\left ( {d\sigma}/{d\Omega}\right)_0$ is the differential cross section with unpolarized particles, $\lambda =\pm 1$ is the photon helicity (we suppose, for simplicity that the photon beam is fully circularly polarized), $t_x$ and $t_z$ are the possible components of the target proton polarization $\vec t$. ${\cal A}_x$ and ${\cal A}_z$ are the T-even asymmetries for the process 
$\vec \gamma+\vec p\to p+V^0$, which are real functions of the photon energy $E_{\gamma}$ and $\cos\theta$, $\theta$ is the V-meson production angle in the reaction CMS. The coordinate system is defined as follows: the $z-$axis is along the photon three-momentum $\vec k$ and the $xz$-plane coincides with the reaction plane, so that the $y-$axis is along the normal to the scattering plane.

The parametrization (\ref{eq:sig}) is very general and holds for any binary process in $\vec \gamma+\vec p$-collisions, due to the P-invariance of the electromagnetic interaction of hadrons.

\begin{figure}
\mbox{\epsfysize=15.cm\leavevmode \epsffile{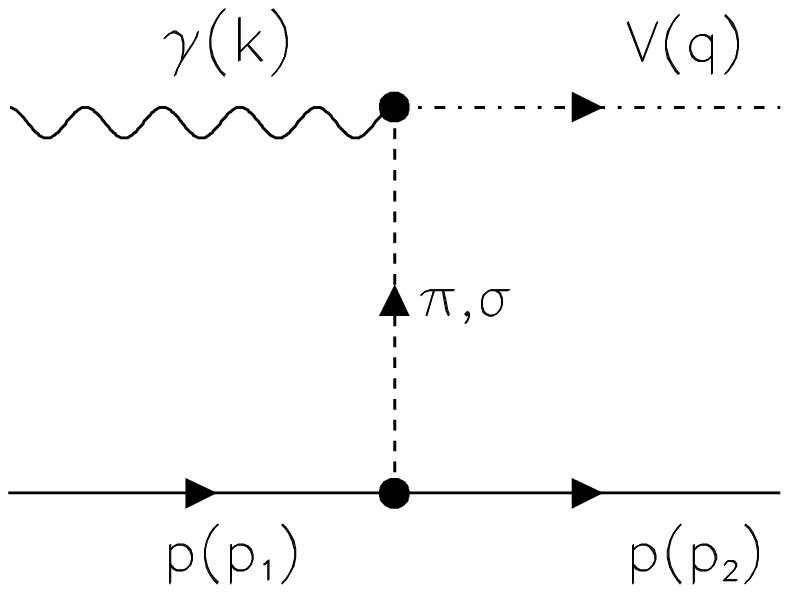}}
\vspace*{.2 truecm}
\caption{Feynman diagram for $\gamma+p\to p+V$, corresponding to the exchange of a scalar ($\sigma$)and a pseudoscalar ($\pi$-meson).}
\label{Fig:fig1}
\end{figure}

Let us calculate these asymmetries, in framework of the $\sigma+\pi$-exchange model, for $\gamma+ p\to p+V$ (Fig. \ref{Fig:fig1}). The Lagrangians for $\pi NN$- and $\sigma NN$- interactions are:
\begin{eqnarray}
& {\cal L}_{\pi NN}&=g_{\pi NN}\overline{N}\gamma_5\vec\tau\cdot\vec\pi N,\nonumber \\
& {\cal L}_{\sigma NN}&=g_{\pi NN}\overline{N} N\sigma,
\label{mdst}
\end{eqnarray}
where $g_{\pi NN}$ is the coupling constant for the pseudoscalar $\pi NN$-interaction and $g_{\sigma NN}$ is the coupling constant of the $\sigma NN$-vertex, which is uniquely defined.

The corresponding electromagnetic interactions with vector mesons are described by the following Lagrangians:
\begin{eqnarray}
& {\cal L}_{V\pi \gamma}&= \displaystyle\frac{e}{2m_V}g_{V\pi \gamma}\epsilon_{\mu\nu\alpha\beta}F_{\mu\nu}
\displaystyle\frac{dV_{\alpha}}{dx_{\beta}},\nonumber \\
& {\cal L}_{V\sigma \gamma}&=\displaystyle\frac{e}{2m_V}g_{V\sigma \gamma}
F_{\mu\nu}(V_{\nu}d_{\mu}\sigma-V_{\mu}d_{\nu}\sigma),
\label{lag}
\end{eqnarray}
where $e$ is the electric charge, so that $e^2/4\pi=\alpha$, $\alpha=1/137$, $m_V$ is the vector meson mass, $F_{\mu\nu}$ is the photon field strength tensor and $V_{\alpha}$ is the vector meson field. In the momentum space the Lagrangians (\ref{lag}) induce the following spin structure for $V\pi\gamma$- and $V\sigma \gamma$- vertices:
\begin{eqnarray}
& {\cal M}_{V\pi\gamma}&=ie\displaystyle\frac{g_{V\pi \gamma}}{m_V}\epsilon_{\mu\nu\alpha\beta}a_{\mu}k_{\nu}U_{\alpha}q_{\beta},
\nonumber \\
& {\cal M}_{V\sigma \gamma}&=e\displaystyle\frac{g_{V\sigma \gamma}}{m_V}(a\cdot Uk\cdot q -a\cdot q U\cdot k),
\label{mat}
\end{eqnarray}
where $a_{\alpha}$ and $k$ ($ U_{\alpha}$ and $q$) are the four vector of polarization and the four momentum of the photon (V-meson), so
$$ a\cdot k=U\cdot q=0.$$

The matrix element, corresponding to $\pi+\sigma$-exchange, in the process 
$\gamma+N\to N +V$, Fig. \ref{Fig:fig1},  can be written as follows:
$${\cal M}={\cal M}_{\pi}+{\cal M}_{\sigma},$$
\begin{eqnarray}
& {\cal M}_{\pi}&=ie
\displaystyle\frac{g_{\pi NN}}{t-m^2_{\pi}}
\displaystyle\frac{g_{V\pi \gamma}}{m_V}
F_{\pi}(t)\epsilon_{\mu\nu\alpha\beta}a_{\mu}k_{\nu}U_{\alpha}q_{\beta}
\overline{u}(p_2)\gamma_5 u(p_1),\nonumber \\
& {\cal M}_{\sigma}&=e
\displaystyle\frac{g_{\sigma NN}}{t-m^2_{\sigma}}
\displaystyle\frac{g_{V\sigma \gamma}}{m_V}F_{\sigma}(t)(a\cdot U k\cdot q-a\cdot q U\cdot k) \overline{u}(p_2) u(p_1),
\label{eq:mat}
\end{eqnarray}
where $t=(p_1-p_2)^2=(k-q)^2$ (the notation of the particle four-momenta is shown in Fig. \ref{Fig:fig1},  $F_{\pi}$ and $F_{\sigma}$ are phenomenological form factors, which insure the correct $t-$dependence of the differential cross section $d\sigma(\gamma p\to pV)/dt$ in the framework of the considered model.

The matrix element (\ref{eq:mat}) satisfies the condition of gauge invariance of the electromagnetic interaction - for any value of the coupling constants, form factors $F_{\pi}$ and $F_{\sigma}$ and in any kinematical condition.

The sum ${\cal M}_{\pi}+{\cal M}_{\sigma}$ generates very particular polarization phenomena. First of all, due to the fact that the amplitudes are real (in contradiction with the unitarity condition) all T-odd polarization phenomena, such as, for example, the analyzing powers (induced by the target polarization) or the components of the polarization of the final protons (in collisions of unpolarized particles) are identically zero. The simplest T-even polarization observable, the $\Sigma_B$-asymmetry, induced by linear polarized photons  also vanishes, due to the specific spin structure of both electromagnetic vertices, $V\pi \gamma
$ and $V\sigma \gamma$, which enter in the considered model with $\pi+\sigma$-exchange. 

The density matrix of the $V$-meson is $\rho_{xx}=\rho_{yy}=1/2$ - with $\sin^2\theta_{\pi}$-dependence  of the decay products in $\rho^0\to\pi^++\pi^-$. These predictions hold for any kinematical condition, for the considered reactions in the framework of $\pi+\sigma$-exchange, independently on the value of the coupling constants and the parametrization of the form factors $F_{\pi}$ and $F_{\sigma}$.

Such predictions on polarization observables, if experimentally verified, constitute a powerful test of the validity of the model, for $\gamma+p\to p+\rho^0$. The T-even polarization observables  ${\cal A}_x$, ${\cal A}_z$, defined in Eq. (\ref{eq:sig}), which are not vanishing, bring information on the details of the model.

The asymmetry ${\cal A}_z$ is related to the Gerasimov-Drell-Hearn sum rule \cite{GDH}, which applies to the process of vector meson photoproduction on nucleons, $\gamma+N\to N+V$. Typically, due to the relatively large threshold, its contribution is not so large \cite{Zh02}. But this asymmetry is considered here from another point of view. 

The angular dependence of the asymmetry ${\cal A}_x$ follows a $\sin\theta$-distribution, as a result of the helicity conservation for the collinear regime.

Using the matrix elements, Eq. (\ref{eq:mat}), one can find the following formulas for the asymmetries ${\cal A}_x$ and ${\cal A}_z$ in terms of the corresponding coupling constants (assuming, for simplicity, $F_{\pi}(t)=F_{\sigma}(t)$):

\begin{equation}
{\cal A}_{x,z}=-2R\lambda\left (\displaystyle\frac{m^2_{\pi}-t}{m^2_{\sigma}-t}\right )N_{x,z}, 
\label{eq:eq6}
\end{equation}
\begin{equation}
DN_z=\displaystyle\frac{m}{2E_ {\gamma}}
\left [ -1-\displaystyle\frac{W^2}{m^2}+\displaystyle\frac{2}{t}m_V^2\right ],
\label{eq:eq7}
\end{equation}
\begin{equation}
DN_x=\displaystyle\frac{W}{2E_{\gamma}}
\sqrt{-\left (1-\displaystyle\frac{t_{min}}{t}\right )
\left (1-\displaystyle\frac{t_{max}}{t}\right )},
\label{eq:eq8}
\end{equation}
\begin{equation}
D=1+R^2\left (\displaystyle\frac{m^2_{\pi}-t}{m^2_{\sigma}-t}\right )^2
~\left (1-\displaystyle\frac{4m^2}{t}\right ),
\label{eq:eq9}
\end{equation}
\begin{equation}
R=\displaystyle\frac{g_{V\sigma \gamma}}{g_{V\pi \gamma}}\displaystyle\frac{g_{\sigma  NN}}{g_{\pi NN}},
\label{eq:eq11}
\end{equation}
where $m$ is the nucleon mass and $t_{^{max}_{min}}=m^2_V-2\omega E\pm 2\omega q$ and 
$$E=\displaystyle\frac{W^2+m^2_V-m^2}{2W},
~\omega=\displaystyle\frac{W^2-m^2}{2W}=E_ {\gamma}\displaystyle\frac{m}{W},~q = \sqrt{E^2-m^2_V}, $$
$E_ {\gamma}$ is the photon energy in the c.m.s. and $W^2=m^2+2E_ {\gamma}m$.

\begin{figure}
\mbox{\epsfysize=15.cm\leavevmode \epsffile{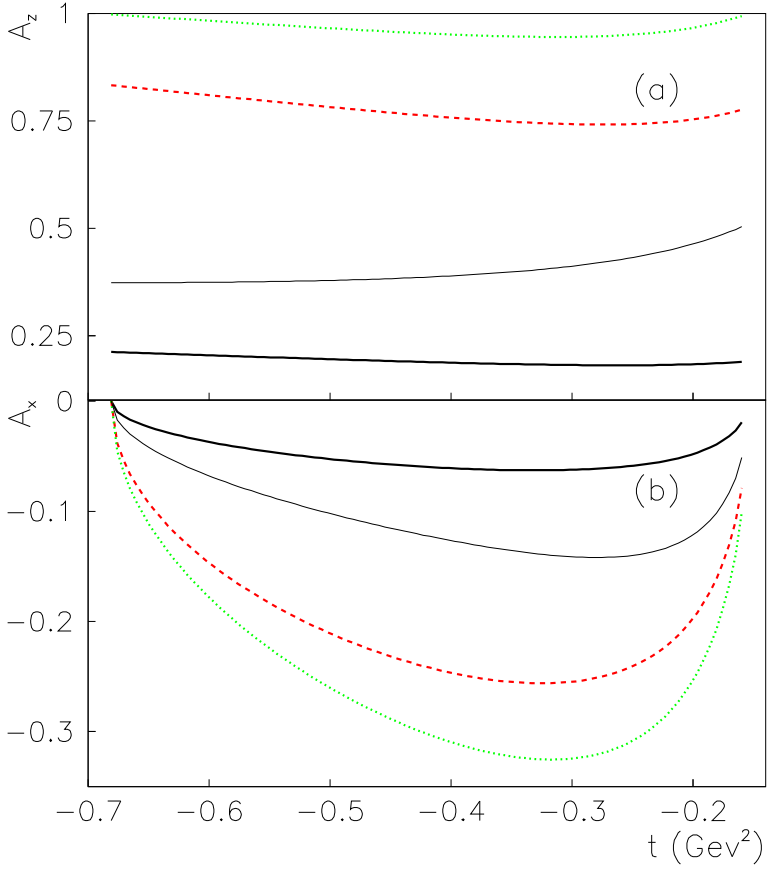}}
\vspace*{.2 truecm}
\caption{$t$-dependence of the asymmetries ${\cal A}_z$ (a) and ${\cal A}_x$ (b) at $E_{\gamma}$=1.2 GeV, for different values of ${\cal R}$ and $m_{\sigma}$:  ${\cal R}$=5 and $m_{\sigma}$=0.5 GeV (thick solid line), ${\cal R}$=5 and $m_{\sigma}$=1 GeV (thin solid line), ${\cal R}$=1 and $m_{\sigma}$=0.5 GeV (dashed line), ${\cal R}$=0.5 and $m_{\sigma}$=0.5 GeV (dotted line):}
\label{Fig:fig2}
\end{figure}

\begin{figure}
\mbox{\epsfysize=15.cm\leavevmode \epsffile{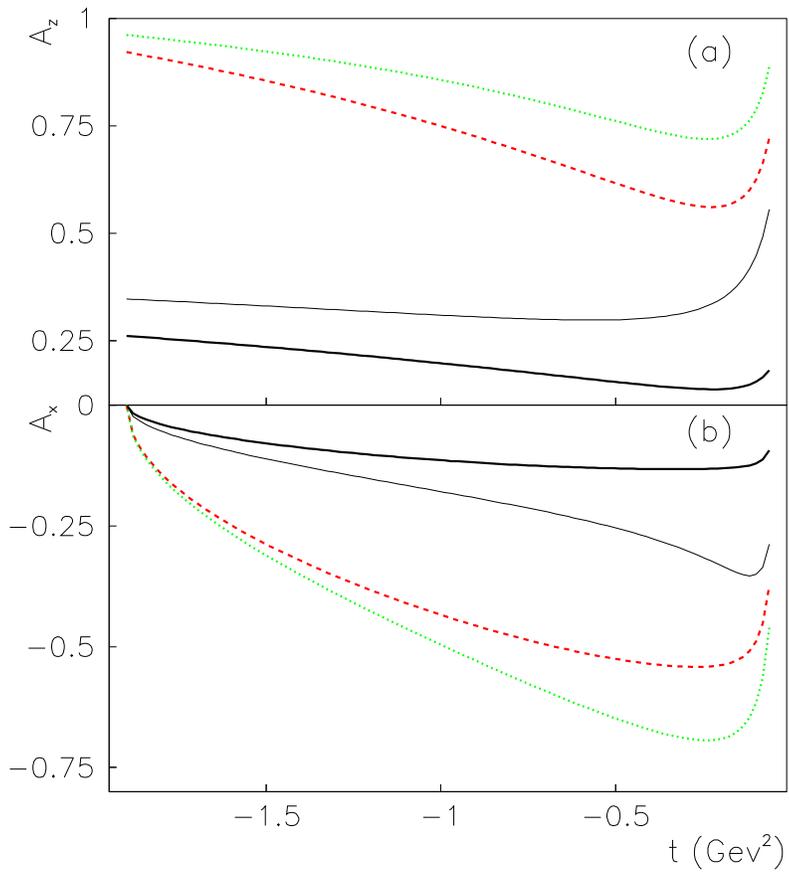}}
\vspace*{.2 truecm}
\caption{Same as Fig. 2, for $E_{\gamma}$=1.8 GeV.}
\label{Fig:fig3}
\end{figure}

At a fixed value of the ratio $R$, the asymmetries ${\cal A}_x$ and ${\cal A}_z$, are functions of  $E_ {\gamma}$  and of the momentum transfer squared, $t$, which varies in the interval: $t_{min}\le t \le t_{max}$.
Figs. \ref{Fig:fig2} and \ref{Fig:fig3} illustrate the sensitivity of these asymmetries for the processes $\gamma+p\to p+\rho^0(\omega)$ to the ratio $R$ of the coupling constants  $g_{V\sigma \gamma}$ and $g_{V\pi \gamma}$ as functions of $t$, for different photon energies (in the near threshold region). One can see that the asymmetry ${\cal A}_z$ is characterized by a weak energy dependence,  and by a relatively weak $t$-dependence, in the considered  $E_{\gamma}$-interval, for a wide interval of ${\cal R}$ and $m_{\sigma}$. Being positive in the considered kinematical region, this asymmetry shows an essential sensitivity to the ratio ${\cal R}$ for any value of $t$ and for the two values of $E_{\gamma}$, $E_{\gamma}=1.2$ and 1.8 GeV.

The asymmetry ${\cal A}_x$ is large (in absolute value) for small values of $t$, where the differential cross section is larger - with a strong energy and $t-$dependence. This asymmetry is also sensitive to the ratio ${\cal R}$ and to $m_{\sigma}$. 

Note that, in the framework of the considered model, $\pi$+$\sigma$-exchange, the relative values of the asymmetries ${\cal A}_x/{\cal A}_z$ can be predicted exactly, without any dependence on $m_{\sigma}$ and on the different coupling constants, being therefore a good test of the validity of the model.

Note that, in a similar way, it is possible to consider also similar polarization phenomena for the inelastic vector meson processes \cite{So00}: 
\begin{equation}
\vec\gamma+\vec p\to N^*({\cal J}^P=1/2^{\pm})+V,
\label{eq:eq12}
\end{equation}
where $N^*({\cal J}^P=1/2^{\pm})$ is a nucleon resonance with spin ${\cal J}$ and parity $P$ equal to  
${\cal J}^P=1/2^{+} $ (the Roper resonance) or ${\cal J}^P=1/2^{-} $  (the N(1535)). Applying the same model of $\pi+\sigma$-exchanges, one can see that the $\sigma$ and $\pi$ contributions to the matrix element of the process (\ref{eq:eq11}) do not interfere in the differential cross section with unpolarized particles. Therefore the $\Sigma_B$-asymmetry vanishes, but with the same the sensitivity to the relative role of $\pi^{-}$ and $\sigma$-exchanges for the double polarization observables ${\cal A}_x$ and ${\cal A}_z$. However, instead of the parameter $R$, one has to define another ratio,  $R^*$,
$$R^*=\displaystyle\frac{g_{V\sigma \gamma}}{g_{V\pi \gamma}}
\displaystyle\frac{g_{\sigma  NN^*}}{g_{\pi NN^*}}.$$
Therefore, the measurement of the two ratios, $R$ and  $R^*$ allows to determine the 
$\sigma  NN^*$ coupling constant:
$$\displaystyle\frac{R^*}{R}=\displaystyle\frac{g_{\sigma  NN^*}}{g_{\pi NN^*}}/
\displaystyle\frac{g_{\sigma  NN}}{g_{\pi NN}}.$$
This ratio is independent on the radiative coupling constants, therefore the ratio 
$ R^*/R$ is of the same order for $\omega$ and $\rho$-photoproduction, if the $\sigma+\pi$-model applies.

Note that the circular polarization of the photon beam results also in non-zero ${\cal P}_x$- and ${\cal P}_z$-components of the final proton polarization.  Generally the observables ${\cal P}_x$ and ${\cal P}_z$ contain new dynamical information in comparison with the asymmetries  ${\cal A}_x$ and ${\cal A}_z$, but it is not the case in the considered model, where they are determined by the same formulas.

\section{Conclusions}
We derived the cross section and the polarization observables for the photoproduction of vector mesons on polarized protons, with a circularly polarized beam, $\vec\gamma+\vec p\to p+V^0$, $V^0=\rho,\omega$ or $\phi$. We applied the $\sigma+\pi$-model in the near threshold region and gave predictions for the polarization observables. The two possible asymmetries ${\cal A}_x$ and ${\cal A}_z$ have been determined through the $\sigma\bigotimes\pi $-interference, with a resulting sensitivity to the sign and the absolute value of the ratio $R$ of the strong and the electromagnetic coupling constants (Eq. \ref{eq:eq11}). Taking into account that the $g_{V\pi \gamma}$ and the $g_{\pi NN}$ coupling constants are well known from other sources, the measurements of the ratios $R_{\omega}$ and $R_{\rho}$ will open the possibility to determine the $g_{V\sigma \gamma}$- couplings. Taking the $SU(3)$ prediction for the ratio $g_{\omega\pi \gamma}/g_{\rho\pi \gamma}\simeq 3$, which is in good agreement with the existing data on the widths of the radiative decays $V\to \pi\gamma$, one can find:
$$ \displaystyle\frac{R_{\rho}}{R_{\omega}}=\displaystyle\frac{g_{\rho\sigma \gamma}}
{g_{\omega\sigma \gamma}}\displaystyle\frac{g_{\omega\pi \gamma}}{g_{\rho\pi \gamma}}\simeq 3\displaystyle\frac{g_{\rho\sigma \gamma}}{g_{\omega\sigma \gamma}}.$$
It is important to note that the ratio $R_{\rho}/R_{\omega}$ does not depend on the 
$g_{\sigma NN}$ coupling constant. The comparison of the obtained values of 
$g_{\rho\sigma \gamma}/g_{\omega\sigma \gamma}$ with this ratio from the decay $\omega(\rho)\to \pi^0+\pi^0+\gamma$, will constitute a good test of the existing theoretical approaches.

Note that the asymmetries ${\cal A}_x$ and ${\cal A}_z$, for the vector meson photoproduction on neutrons $\vec\gamma+\vec n\to n+V^0$ have the same absolute values, but differ in sign from the corresponding asymmetries on polarized proton target. This is another possible test of the validity of the $\sigma+\pi$-model as any additional contribution to the matrix element of the considered process violates this relation.


\end{document}